\documentclass[aps,pra,twocolumn,10pt,superscriptaddress, showpacs,amsmath,amssymb,nofootinbib ]{revtex4-1}

\usepackage{graphicx}
\usepackage{hyperref}
\usepackage{array}
\usepackage{enumitem}

\graphicspath{{./figures/}}

\newcommand{\ohm}{$\Omega$}

\begin{document}

\title{Redesigning a junior-level electronics course \\
to support engagement in scientific practices}

\author{H. J. Lewandowski}

\affiliation{Department of Physics, University of Colorado, Boulder, CO 80309, USA}
\affiliation{JILA, National Institute of Standards and Technology and University of Colorado, Boulder, CO, 80309, USA}

\author{Noah Finkelstein}
	\affiliation{Department of Physics, University of Colorado, Boulder, CO 80309, USA}

\begin{abstract}
Building on successful work on studying and transforming our senior-level Advanced Lab course, we have transformed our junior-level Electronics course to engage students in a variety of authentic scientific practices, including constructing, testing, and refining models of canonical measurement tools and analog circuits. We describe our approach to the transformation, provide a framework for incorporating authentic scientific practices, and present initial outcomes from the project. As part of the broader assessment of these course transformations, we examine one course learning outcome: development of the ability to model measurement systems. We demonstrate that in the transformed course students demonstrate greater likelihood of identifying discrepancies between the measurement and the model and significantly greater tendencies to refine their models to reconcile with the measurement.
\end{abstract}

\maketitle

\section{Introduction}
\vspace{-0.4cm}

Recently, there has been increased interest in improving undergraduate physics laboratory instruction at both the introductory\cite{etkina,pcast} and advanced-lab level\cite{ajplew}. In particular, the national community of lab instructors and physics education researchers are beginning to discuss and agree on explicit desired learning outcomes for physics lab courses. In 2014, the American Association of Physics Teachers endorsed the \textit{Recommendations for the Physics Laboratory Curriculum}, which is an extensive list of learning outcomes for all levels of lab courses \cite{aapt}. This document parallels the University of Colorado's (CU) own learning goals document developed to guide transformation of the Advanced Lab course at CU \cite{ajplew}.

CU's goals can be grouped into four main categories of scientific practices: modeling, design, communication, and technical lab skills. Each of these categories has many detailed sub-goals\cite{ajplew}. It is clear that all of these learning objectives can not be met with one three-credit lab course. After several years of working on transforming our Advanced Lab course, we began a process to restructure our junior-level Electronics course to also align with these goals. 

There are many advantages to engaging students in these scientific practices in an upper-division \emph{electronics} course. First, unlike typical advanced lab courses, the equipment is relatively low cost, and thus enough can be purchased to have all of the students complete the same lab activities each week. This enables the course activities to be adequately scaffolded to introduce students to scientific practices through guided labs and move towards more open-ended projects with much less direct guidance. Second, sophistication of the equipment and components is at the right level to allow students to construct, test, and refine models of the systems. Advanced lab equipment can be extremely complex (e.g., STM, laser cooling and trapping apparatus) and thus hinder students' ability to construct models of the experiments. Finally, concentrating on scientific practices in the junior-level course reduces some of the burden to meet all of the learning goals in the Advanced Lab course and prepares students earlier in their careers to participate in undergraduate research. Ultimately, these scientific practices should be taught in the lower-division labs as well. 
  
 Here, we describe the redesign of an electronics course to engage students in several scientific practices, and how explicit prompts can help students begin to develop and refine models. We show that students can engage in the sophisticated activity of modeling measurement systems, and do not initially engage in model refinement without prompting. 
 
 \vspace{-0.5cm}
  
\section{Overall Redesign of Course}
\vspace{-0.2cm}

	CU's junior-level Electronics course is required of all physics majors and enrolls about 50 students every semester. Each section of the course has no more than 20 students and it taught by regular physics faculty members. The course consists of one three-hour lab section and two one-hour lectures each week. Students have card access to the lab room nearly 24--7 to be able to complete the lab activities.  The students work in pairs on guided lab exercises during the first 10 weeks of the semester. The final 5 weeks are dedicated to student-proposed projects. The course covers mostly typical analog electronics (e.g., op-amps, filters, transistors) with two weeks devoted to digital gates and Arduinos. 

\begin{figure}[h]
  \begin{center}
  
  \includegraphics[scale=1]{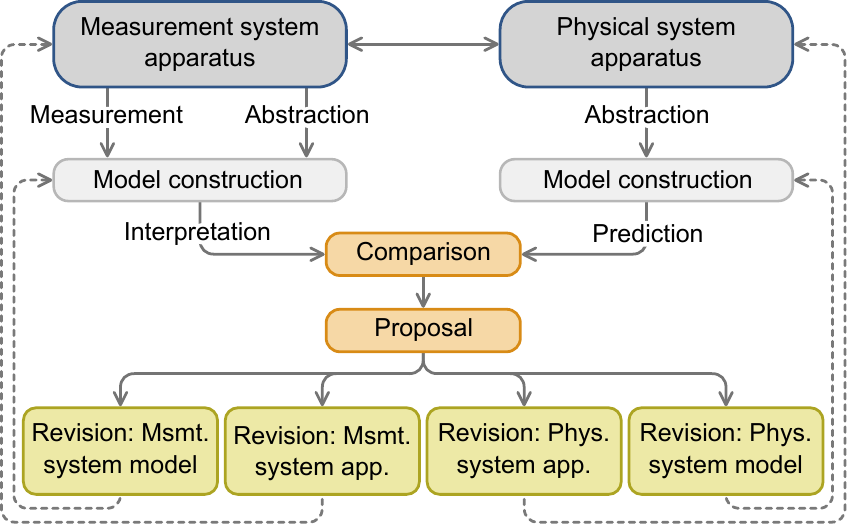}
  \end{center}
  \caption{Experimental modeling framework. The framework describes the iterative process of constructing models of the measurement and physical systems, comparing measurements to predictions, proposing refinements to address discrepancies,
and making revisions to models and/or apparatuses. This figure is a simplified version of the full framework presented by Zwickl et al. in Ref. \cite{specialBZ}.}
\label{framework}
\end{figure}

	The overall course transformation was designed to address several goals including: (1) making the experience more authentic and aligned with current practices of experimental physicists working with electronics \cite{pollard}, (2) including more design and application activities in the guided-lab portion of the course, and (3) preparing students for their projects and making them more accountable during this portion of the course. There were also three main items that we stated were not goals for the course including: (1) formal error analysis, (2) understanding a historical perspective on electronics, and (3) understanding the details of relevant solid-state physics. We modified the course in many ways to address these goals. 
	
\textbf{(1) Enhanced Student Preparation for labs:}

a.	Prelab assignments were redesigned to have students review the entire lab guide, review theory, and construct mathematical models to make predictions (e.g., plots in Mathematica) that could be tested in lab by measurement.

b.	Laboratory Skill Activities were introduced. Two tutorial activities, with online screencasts, were produced for students to get up to speed on the use of Mathematica in a lab environment.

\textbf{(2) Emphasized collecting / analyzing /reporting on experiments during class time:}

a.	The use of a laboratory notebook was enhanced. Students were coached (provided goals, approach, and exemplar versions) on how scientific lab notebooks are kept.  The notebook served as the primary form of communication with the faculty. It was used to record the specific activities, measurements, models, and analysis of the lab. Students were encouraged to print out plots and tape them into the lab notebook. The goal was to have students do most of the analysis and plotting in the lab.

b.	Analysis of data during class time was encouraged. By preparing for the lab with the prelabs assignments, students could directly compare measurements with predictions from theory during the class time. This facilitated analysis, debugging, communication with faculty, and emphasized the modeling objectives of the course.

c.	Active participation in lab activities was encouraged. By placing some portion of student grades on participation in laboratory, we encouraged students to be actively engaged during their assigned laboratory section (for both the guided labs and the projects.)  Such emphasis promoted community, engagement, and opportunities for interaction with faculty.  (Students could still maintain autonomous schedules to finish their laboratory work.) 

d.	Traditional lab reports were eliminated.  To reduce workload, emphasize authentic scientific practice, and focus faculty interactions on productive, formative feedback, the enhanced lab notebook replaced the traditional lab reports.  

\textbf{(3) Increased accountability and engagement in student-designed projects:}

a.	The project proposal phase was extended to include three stages. The stages included developing: i) an idea or research question, ii) a coarse description of the project, and iii) detailed outline of the project, which included circuit diagrams, weekly plans, goals, and predicted challenges. 

b.	Weekly project updates were included. During class time, students were asked to present their work to the lab instructor.  This provided practice for the final presentation and an opportunity to get corrective feedback on the overall project objectives (rather than the specifics of a circuit, which happens in the more informal interactions with faculty).

c.	Lab Notebooks were still collected and provided more practice at keeping a scientific notebook and organizing ideas / data for the final project presentation and write-up.

 \vspace{-0.4cm}

\section{Study Design and Methods}

All of the changes described in Section II impacted student outcomes from the course, but, for the rest of this paper, we concentrate on a core goal of engaging students in the authentic practice of modeling measurement systems. Modeling measurement systems is a integral piece of the general practice of modeling, where modeling includes constructing, testing, and refining models. Previously, we developed a framework that describes the modeling process in experimental physics (Fig. \ref{framework})\cite{specialBZ}. Using this framework as a guide, and taking a scaffolding approach to developing student expertise, we rewrote all of the lab guides to encourage students to employ the modeling process to understand their electrical circuits. The lab guides for the labs earlier in the semester were very explicit about the process of modeling. The later guides faded in the explicit guidance and allowed students more freedom about when and how to model their systems. 

To study the impact of the curricular changes on students' modeling skills, we examine an activity from the lab guide for the second week of class from both the traditional offering of the course and the transformed offering of the course. As this was the first time in their undergraduate curriculum students were introduced to the idea of modeling (in each format of the course), we wanted to answer the research question ``does explicit prompting of modeling activities in a lab guide result in students' application of modeling to their experiments?'' An obvious followup question of ``do students engage in modeling activities after the explicit prompts have been removed?'' is a focus of future studies. 

\begin{figure}[h]
  \begin{center}
  \includegraphics[scale=0.35]{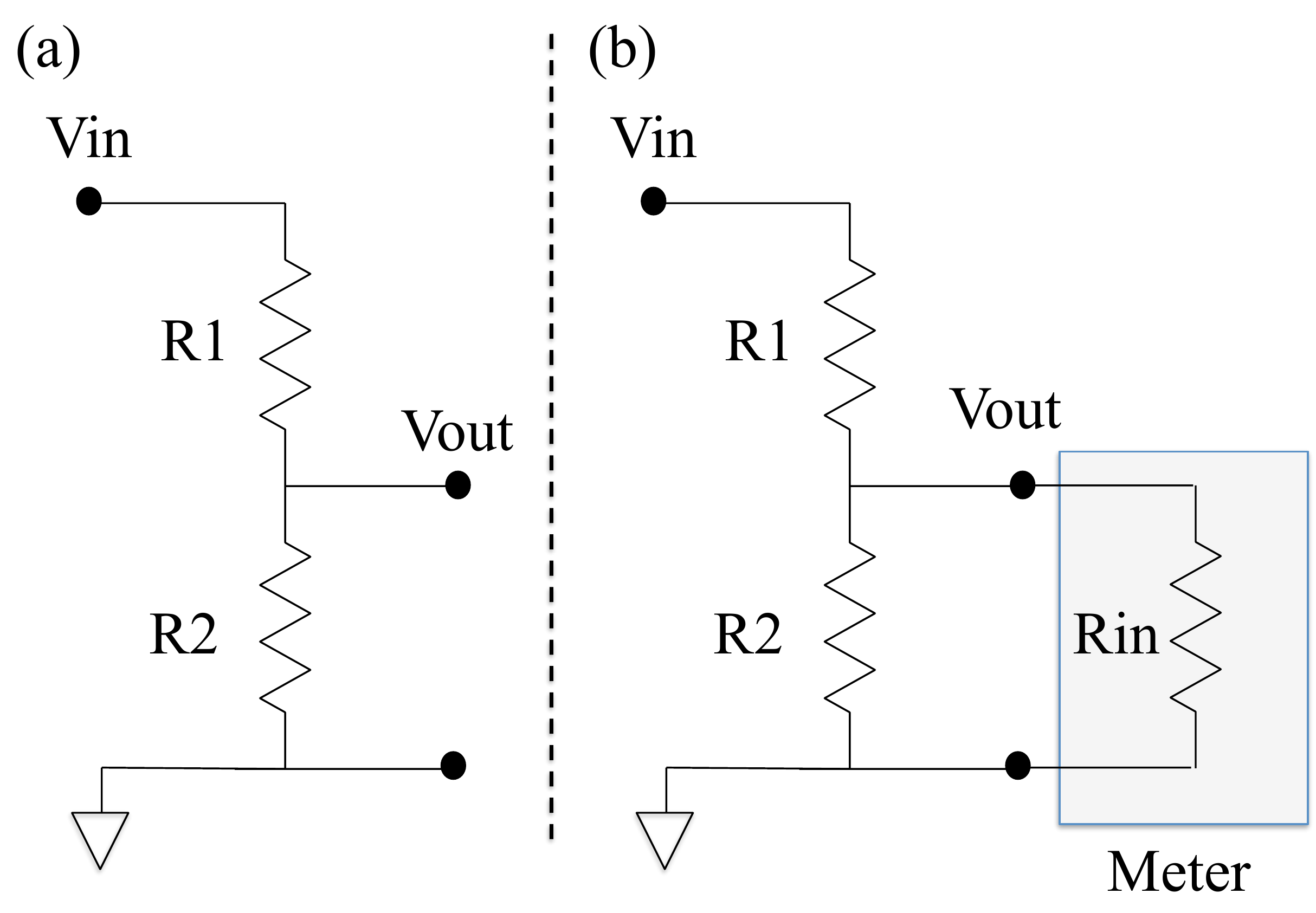}
  \end{center}
  \caption{Circuit diagrams of voltage dividers. (a) Diagram given to students to build as part of the lab. (b) Revised circuit diagram that students developed. This includes modeling the measurement tool (meter) as a parallel resistance.}
  \label{diagram}
\end{figure}

In each of the two conditions, the lead instructor was the same (lead author) and the courses were run one year apart.  Student demographics (typically third-year physics majors) are nearly identical in the two data sets, predominantly ($>$90\%) male, and dominantly ($>$90\%) white.  The students are asked to conduct the same experiment, building three different voltage dividers (Fig. \ref{diagram} a) using nominally identical resistors for $R_1$ and $R_2$ of 1 k\ohm, 1 M\ohm, and 10 M\ohm. They are instructed in both guides to apply a dc voltage to $V_{in}$, measure $V_{out}$ using both a digital multimeter (DMM) and an oscilloscope, and finally compare the measurements with their predictions. 

The expert will recognize that the input impedance of the measurement device ($R_{in}^{DMM}$ = 10 M\ohm, $R_{in}^{scope}$ = 1 M\ohm)  will begin to affect the measured voltage when the impedance becomes comparable to the $R_2$. For this situation, a new model must be used to correctly predict the output voltage. The refined model must have a resistor parallel to $R_2$ to represent the input impedance of the measurement device (Fig. 2b). Alternatively, one could use the new model and the voltage measured to determine the value of the input impedance of the device.
 
 After students measured the output voltages, the traditional lab guide prompts them to compare the measurements to their predictions.
 
\begin{quote}
\emph{Compare the voltages you expected to the voltages you measured. What does this tell you about the input impedance of your instruments?}
\end{quote}

The instructors' desired result is for students to use the measured voltages and a model that includes the measurement device to determine a value for the input impedance. In this case, the students are not given any direct prompting in the lab guide, although they may get direction from the instructor during the lab section.

In the transformed lab guide, the desired modeling steps are explicit prompts (below). The students are first asked to determine if their measurements and predictions agree. This is represented as the \emph{Comparison} step in the Modeling Framework (Fig.\ref{framework}). If they determine the agreement is not acceptable, they are prompted to make a \emph{Revision} of their model. The refinement includes two common representations: a circuit diagram and a mathematical equation relating the input to the output. Finally, they are directed to use the new model to make a \emph{Prediction} of the input impedance of the device. 
 
\begin{quote}\emph{Compare the voltages you predicted to the voltages you measured. Does your model of the voltage divider agree with each of your measurements? Explicitly record what criteria you used to determine whether or not the model and measurements agreed. }\newline

\emph{\indent Complete this step only if your model and measurements did not agree. If your model and measurements did not agree, you will have to either refine your model or your experiment. Let’s start by refining your model. Consider the input resistance of your measurement device. Draw a circuit diagram that includes that resistance.  HINT: You already worked with this circuit model in your prelab. Derive an expression for the output voltage now including the unknown measurement device resistance. Use this new model to determine the input resistance of measurement device.}
\end{quote}

To determine the outcomes from the two different prompts, we coded student write-ups of the lab for evidence of modeling for two semesters, one before the transformation (N = 47) and one after the transformation had been implemented for one year (N = 44).  In the traditional course, we used the formal lab reports and lab notebooks for the transformed course. In each case, we evaluated student work for the course materials that were graded and returned to the students.  It is worthy to note that the lab reports were significantly longer (typically 6-10 pages) and more formal than the lab notebooks (hand-written and computer modeling of 3-5 pp). On the other hand, the emphasis in the transformed course was that the laboratory notebooks would serve as authentic scientific artifacts, with the goal of resembling the notebooks of practicing scientists. 

 \vspace{-0.4cm}
 
\section{Results and Discussion}

The notebooks and reports were both coded with the scheme listed below. We recorded whether or not each student included the following items in their notebook/report when describing the 1 and 10 M\ohm~voltage dividers. Half of the artifacts were coded independently by a second researcher in the group, which resulted in 100\% agreement on the codes presented in Table \ref{table}.

\begin{itemize}\itemsep0pt \parskip0pt \parsep0pt

 \item noted the input impedance could account for the prediction/measurement discrepancy
 \item noted the meter could be modeled as a parallel resistor
 \item included a circuit diagram with $R_{in}$ included
 \item included a new mathematical model
 \item made a new prediction of $V_{out}$
 \item made a prediction of $R_{in}$
 
\end{itemize}
\vspace{-.3cm}

It is worthy to note that there are instances when students in less heavily guided curricula engage more significantly in exploration and identify more potential outcomes of an intellectual space \cite{chamberlin}.  However, because students have had little explicit exposure to modeling and refining models, we hypothesized that the explicit scaffolding of the modeling process would result in greater application of modeling and refining of models of measurement systems in this case.

\begin {table}[h]

\begin{tabular}{ l | c | c }

 \hline		
 Semester & Identified discrepancy & Refined Model \\
 \hline
 Trad’l (N = 47) & 83\% & 27\% \\
 Transfrm (N = 44) & 100\% & 100\% \\
 \hline  

\end{tabular}
\caption {Engagement in modeling in the traditional and transformed classes. The table shows the percentage of students who correctly identified the input impedance of the meter as the source of the discrepancy and the percentage that refined the model to include this impedance to make a prediction. The two samples (Tradl and Transformed) are different for each of the two cases: Idenitfy Discrepancy p $<$ 0.01 and Refine Model p $<$ 0.001 by measure of binomial sampling.}
\label{table}
\end {table}
 \vspace{-0.2cm}

Selected results from coding the lab notebooks and reports are shown in Table \ref{table} for both traditional and transformed courses. The second column lists the percentage of student who identified that the discrepancy was likely due to the finite input impedance of the scope or DMM. The third column list the percentage of students who go on to refine the model and make a prediction. This percentage does not include students who draw a new diagram or derive a new mathematical model without also making a prediction of either $V_{out}$ or $R_{in}$. 

It is clear in this instance that the transformed course materials engages students in more significant comparison between model and measurement, even though both curricula prompted students to make such a comparison. In the traditional environment, this comparison is left as an extension to the cue ``what does this tell you...'' In the transformed environment, the cue is explicit``Does your model .. agree with ...your measurement?''  More substantially, the students in the transformed course are nearly four times as likely as their peers to refine their measurement model. Despite our students having come through transformed theory courses in the prior sequences, they do not resolve discrepancies between model (theory) and measurement, even if they are aware of them. With explicit prompting students universally engage in this expert-like scientific practice. 

Of course, there may be concern that students are being taught merely to be following explicit prompts. Future studies will focus on students capacities for internalizing these sophisticated experimental practices.  Preliminary indication it that students in this transformed environment do just that in later weeks of the course. Both in the guided laboratories and in the student-directed projects, students engage in comparison of models and experiment, and engage in refining either the model or the apparatus. Student work from the course using traditional curricula, however, does not show this evolution in later weeks.   


We suspect these habits of mind, the sophisticated engagement in scientific measurement and modeling, are not simply a matter of revised laboratory prompts, but rather the holistic enterprise of the curricular transformation as described in Section II.  By shifting student preparation for laboratory participation, by coordinating the laboratory class time to authentic forms of experimentation and design, and by having students use authentic tools (material and intellectual) of scientists, we believe students are more likely to engage in productive experimental scientific practices.  Future work will address this hypothesis by expanding on the preliminary outcomes demonstrated here.

\vspace{-0.6cm}
 
\acknowledgments  We thank Bethany Wilcox for coding for inter-rater reliability. This work was supported by NSF grant DUE-1323101.

\bibliography{refs3}

\end{document}